\documentclass[twocolumn,aps,floats,floatfix,superscriptaddress,pra,english]{revtex4}
\usepackage{amssymb}
\usepackage{graphicx}
\usepackage{amsmath}
\usepackage{epsfig}
\usepackage{ulem}

\usepackage[latin9]{inputenc}
\usepackage{array}
\usepackage{multirow}
\usepackage{color}
\usepackage{esint}
\usepackage{bm}
\usepackage{color}
\usepackage{bbm}
\usepackage{hyperref}
\usepackage{babel}
\usepackage{epstopdf}
\usepackage{mathtools}
\usepackage{soul}
\usepackage[dvipsnames]{xcolor}
\usepackage{tikz}
\begin{document}

\setcounter{equation}{0} \setcounter{figure}{0}
\setcounter{table}{0} \setcounter{page}{1} \makeatletter

\global\long\def\id{\mathbbm{1}}
\global\long\def\ui{\mathbbm{i}}
\global\long\def\ud{\mathrm{d}}

\title{Efficient production of a narrow-line erbium magneto-optical trap \\with two-stage slowing}

\author{Bojeong Seo}
\thanks{These authors contributed equally to this work.}
\affiliation{Department of Physics, The Hong Kong University of Science and Technology,\\ Clear Water Bay, Kowloon, Hong Kong, China}

\author{Peng Chen}
\thanks{These authors contributed equally to this work.}
\email{pengchen@ust.hk}
\affiliation{Department of Physics, The Hong Kong University of Science and Technology,\\ Clear Water Bay, Kowloon, Hong Kong, China}

\author{Ziting Chen}
\affiliation{Department of Physics, The Hong Kong University of Science and Technology,\\ Clear Water Bay, Kowloon, Hong Kong, China}

\author{Weijun Yuan}
\affiliation{Department of Physics, The Hong Kong University of Science and Technology,\\ Clear Water Bay, Kowloon, Hong Kong, China}

\author{Mingchen Huang}
\affiliation{Department of Physics, The Hong Kong University of Science and Technology,\\ Clear Water Bay, Kowloon, Hong Kong, China}

\author{Shengwang Du}
\affiliation{Department of Physics, The Hong Kong University of Science and Technology,\\ Clear Water Bay, Kowloon, Hong Kong, China}

\author{Gyu-Boong Jo}
\email{gbjo@ust.hk}
\affiliation{Department of Physics, The Hong Kong University of Science and Technology,\\ Clear Water Bay, Kowloon, Hong Kong, China}

\date{\today}
\begin{abstract}
We describe an experimental setup for producing a large cold erbium (Er) sample in a narrow-line magneto-optical trap (MOT) in a simple and efficient way. We implement a pair of angled slowing beams with respect to the Zeeman slower axis, and further slow down atoms exiting from the Zeeman slower. The second-stage slowing beams enable the narrow-line MOT to trap atoms exiting from the Zeeman slower with higher velocity. This scheme is particularly useful when the Zeeman slower is at low optical power without the conventional transverse cooling between an oven and a Zeeman slower, in which case we significantly improve the loading efficiency into the MOT and are able to trap more than $10^8$ atoms in the narrow-line MOT of $^{166}$Er. This work highlights our implementation, which greatly simplifies laser cooling and trapping of Er atoms and also should benefit other similar elements.
\end{abstract}

\maketitle
% \tableofcontents
\newpage

%\section{}
%\subsection{}

\section{Introduction}
%paragraph*{\bf Introduction}
Ultracold magnetic atoms (e.g. dysprosium and erbium isotopes)  have offered a tunable atomic system where  unprecedented quantum states can be explored. Various remarkable many-body phenomena have been observed in those dipolar quantum gases based on the dipole-dipole interaction, including quantum droplet states~\cite{Schmitt2016,Ferrier-barbut2016,Chomaz2016}, roton excitations~\cite{Chomaz:2018hs,Petter2018}, the extended Hubbard model~\cite{Baier:2016ga} and supersolid phases~\cite{Tanzi2019,Guo2019,Natale2019}. For cooling those atoms, a narrow-line magneto-optical trap (MOT) has popularly been employed using a narrow optical transition with sub-MHz linewidth~\cite{Kuwamoto:1999tb,Xu2003,McClelland2006}. This cooling scheme has the advantage of allowing to pre-cool trapped atoms to a lower Doppler temperature limit, in the $\mu$K or sub-$\mu$K regime, and has been successfully employed for cooling atoms that contain transitions with relatively narrow natural linewidths~\cite{Grunert2002,Yang2015,Bennetts2017,Nosske2017}.  

However, lanthanide elements with their high melting point are still rather difficult to laser cool and trap, although several research groups have successfully realized quantum degenerate gases with Dy~\cite{Lu:2011hl} and Er~\cite{Aikawa:2012ic}. Inspired by earlier work~\cite{Kuwamoto:1999tb}, lanthanide atoms, decelerated in a Zeeman slower using a broad optical transition, are loaded into a narrow-line MOT, but it still remains challenging to fully optimize the transfer of slowed atoms from the Zeeman slower into the narrow-line MOT due to the small capture velocity of the MOT. Moreover, the slowed atoms below this capture velocity often have such high transverse velocities that they miss the MOT capture volume, which further limits the effective atomic flux into the MOT region.  Several methods have been implemented to increase the effective capture velocity of the MOT by combining broad and narrow-line MOTs in a temporal~\cite{Katori:1999jj} or spatial manner~\cite{Lee:2015dg}. Here, we describe a simple and efficient method to increase the loading efficiency into the narrow-line MOT by adding second-stage slowing beams. This method results in a large $^{166}$Er atom number MOT at a temperature $<$20~$\mu$K (after the compression of the MOT), similar to the recent realizations in Yb~\cite{swing:2018bt} and Dy~\cite{Lunden:2019ty} experiments. 

In Er experiments, the large mean velocity of  atoms necessitates relatively high slower beam intensity to supply enough atomic flux into the MOT. However, the high-intensity slower light often induces optical pumping to dark states and/or a radiative force disturbing the weak narrow-line MOT. In the previous works with Er, these effects were avoided by displacing the MOT during the loading stage~\cite{Frisch:2012bb,Ulitzsch:2017hb}. This method, however, may require a large volume vacuum chamber for MOT loading, which limits broad applications, for example, high-resolution optical microscopy. Here, we present an alternative approach for achieving a large sample of cold Er atoms in the narrow-line MOT without such displacement. We add a pair of near-resonant slowing beams between the exit of the Zeeman slower and MOT, which enhances the MOT loading rate by more than an order of magnitude compared to a fully re-optimized system without the angled slower beams. We are able to trap $1.1\times 10^8$ $^{166}$Er atoms in the MOT using angled slowing beams. A key feature of our apparatus is the efficient production of a cold Er sample without using a transverse cooling stage.%, allowing us to operate a external-cavity diode laser setup without a frequency doubling stage for 401~nm.

%XXXXA narrow-line MOT has popularly been employed to pre-cool atoms before achieving a degenerate quantum gas. It is, for example, widely adapted in cooling alkaline-earth-like or lanthanide atoms that contain a transition with relatively narrow natural linewidth in their electronic structure. Nevertheless, it remains subtle to trap a sufficiently large number of atoms in a narrow-line MOT, particularly for erbium  isotopes with the high melting point.XXXX

%We use a two-stage Zeeman slowing to dramatically increase the atoms number in the erbium narrow-line MOT. With moderate cooling laser power, we can load more than $10^8$ atoms in a few seconds. The method  is very simple to implement and does not require extraordinary resource. 

\begin{figure*}
\includegraphics[width=0.8\linewidth]{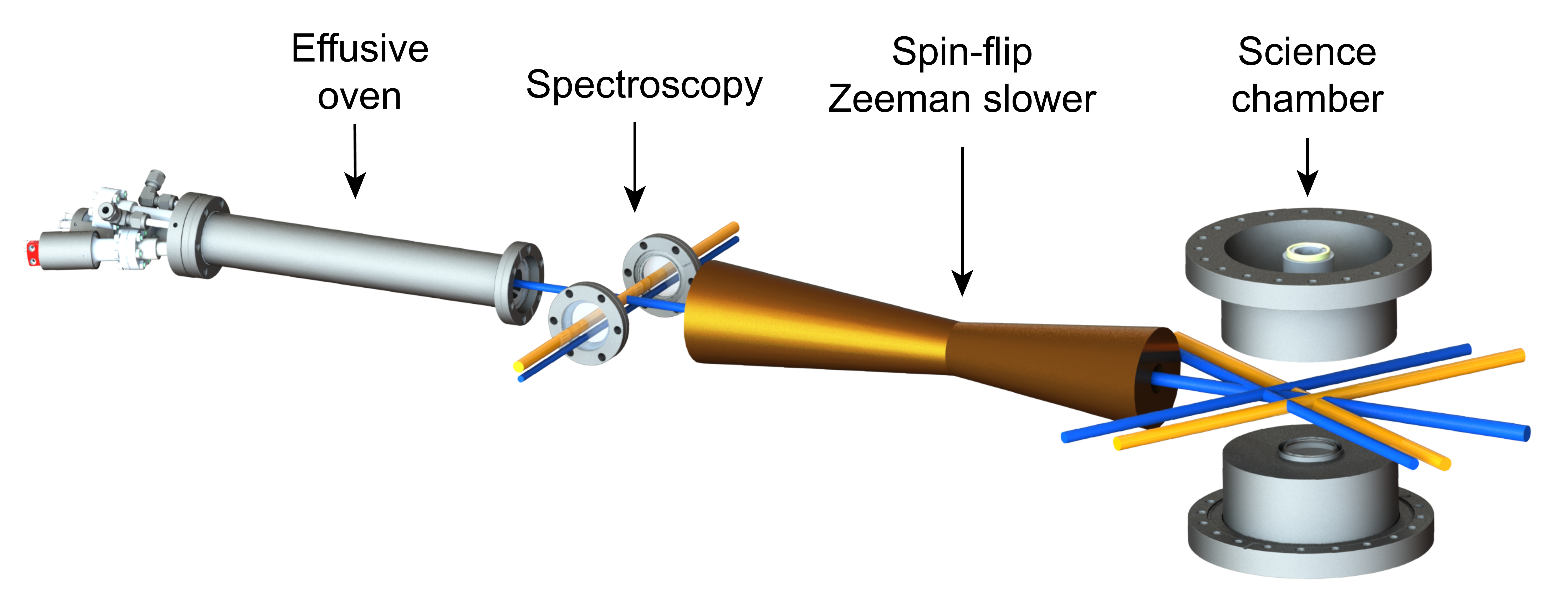}
\caption{ \textbf{Schematic of experimental apparatus for cooling and trapping Er atoms.}  An oven at 1200$^{\circ}$C effuses an Er atomic beam into the intermediate chamber where laser spectroscopy is performed for the 401 nm (blue) and 583 nm (yellow) transitions. The 400 mm-long spin-flip Zeeman slower decelerates atoms and is followed by the second-stage angled slowing beams. Two recessed viewports are used in the science chamber. The 400~mm-long spin-flip Zeeman slower decelerates atoms followed by the second-stage angled slowing beams. Two recessed viewports are used in the science chamber.}
	\label{fig1}
\end{figure*}

%\paragraph*{\bf Effusive oven} 
\section{Experimental setup}
 The design of the experimental appartus is shown in Fig.~\ref{fig1}. The apparatus consists of an effusive atomic oven, an intermediate chamber for laser spectroscopy, a Zeeman slower and a science chamber. A high temperature effusion cell at 1200$^{\circ}$C generates an atomic beam with a mean velocity of 500~m/s from a 4.5~cm crucible containing 25~g of Er pieces (Alfa Aesar 99.9$\%$). The atomic beam is then collimated by a 13~mm-diameter nozzle that consists of 22 pieces of 5~cm  micro-capillary tubes with an inner diameter of 1.4~mm. The nozzle is kept at 1250$^{\circ}$C. Additional 4.5~cm long differential pumping tubes with an inner diameter of 7~mm are located before and after the intermediate chamber. The pressure of the oven and intermediate chamber are $4\times 10^{-11}$ and $1\times 10^{-10}$ mbar, respectively, maintained by ion getter pumps (Gamma vacuum, 45~l/s). The pressure of the science chamber is $5\times 10^{-11}$ mbar.

%\paragraph*{\bf Lasers and frequency stabilization} 
 For laser cooling and trapping Er atoms, two optical transitions at 401~nm and 583~nm are used in this work. The broad transition ($4f^{12}6s^2$ $^3H_6\to$ $4f^{12}  (^3H_6)6s6p(^1P_1)$) at 401~nm with a natural linewidth of $\Gamma_{401}$=$2\pi\times$29.7~MHz is used for the Zeeman slower, angled slower beams and absorption  imaging, while a MOT is formed with the narrow transition ($4f^{12}6s^2$ $^3H_6\to$ $4f^{12}  (^3H_6)6s6p(^3P_1)$) at 583~nm with $\Gamma_{583}$=$2\pi\times$190~kHz. The 401~nm laser light is generated by an external-cavity diode laser (Toptica DL-pro 120 mW output power) as well as a homemade injection-locking setup. The 583 nm laser light is generated by a Toptica TA-SHG system (740~mW output power). 

To stabilize laser frequencies in a simple way, we perform laser spectroscopy with a collimated atomic beam in the intermediate chamber. The 401nm laser is frequency stabilized by standard fluorescence spectroscopy using a 2~mW probe beam with a diameter of $0.8$~mm.  This probe beam is aligned to the edge of the atomic beam without affecting a loading efficiency into the MOT. We stabilize the 583~nm laser using saturation fluorescence spectroscopy, where the fluorescence signal is obtained from a 15~mm-diameter probe  beam with 10~mW by pre-amplifying the photodiode current.  We note that laser spectroscopy does not reduce the atomic flux into the Zeeman slower.

%\paragraph*{\bf  Zeeman slower} 
 Er atoms are slowed by a spin-flip Zeeman slower using the broad optical transition at 401~nm. In contrast to alkali atoms, Er requires relatively high slower laser power due to the large saturation intensity $I_{s,401}=56$~mW/cm$^2$ and the high atomic beam velocity originating from the high oven temperatures (around 1200$^{\circ}$C) required to produce sufficient atomic flux~\cite{melting}. However, a high intensity Zeeman slower beam may cause off-resonant pumping and/or atomic loss in the Er MOT~\cite{Frisch:2012bb,Ulitzsch:2017hb}. This detrimental effect has been alleviated by displacing the MOT below the center of the MOT quadrupole field~\cite{Frisch:2012bb,Ulitzsch:2017hb}.  For this, the detuning of the MOT is increased so that atoms are displaced by gravitational sag. In our work, however, the trapped atoms are difficult to displace due to the recessed viewport being only $\sim$6.5~mm away from the center of the chamber. To overcome this limitation, our Zeeman slower is designed to operate at relatively low intensity and  minimizes the above-mentioned detrimental effect.

%For ytterbium isotopes with a relatively large natural linewidth and the corresponding saturation intensity $I_{sat}=59$~mW/cm$^2$, it is challenging to keep the deceleration ratio $\eta$ as large as $\sim$0.5, as the available laser power at the wavelength of 399~nm is limited. Previous works (for example, see Ref.~\cite{Hansen:2013gp}) employing a Zeeman slower compromised the deceleration factor with relatively high slower laser power up to 100~mW. In that case, the length of the Zeeman slower, proportional to $\frac{v_c^2}{a_{max}\eta}$ where $v_c$ is the capture velocity of the MOT, is set to around 30~cm. 

Now, we discuss the Zeeman slower used in our work. The 400~mm-long spin-flip Zeeman slower consists of five separate parts of coils wrapped around a thin brass tube with a $1"$ outer diameter~\cite{coils}. The coils are wound using insulated square hollowed copper tube. The square tube has an edge length of $1/8"$ and an inner hole diameter of $1/16"$, and is water-cooled~\cite{Song:2016fz}. %It provides a slowing efficiency $\eta=a_{zs}/a_{max}$=0.23 where $a_{zs}$ is deceleration of the Zeeman slower and $a_{max}$ is maximum deceleration achievable,
 The 22~mW slower light is red-detuned by 540~MHz from the singlet transition ($4f^{12}6s^2$ $^3H_6\to$ $4f^{12} (^3H_6)6s6p(^1P_1)$) corresponding to 18.2~$\Gamma_{401}$, resulting in a capture velocity of $~\sim$200~m/s. The residual magnetic field (around 3~G) at the MOT position is compensated by the compensation coil  between the Zeeman slower and the MOT. We note that the Zeeman slower can be operated by a single external-cavity diode laser.

\begin{figure}%[!htb]
	\includegraphics[width=0.95\linewidth]{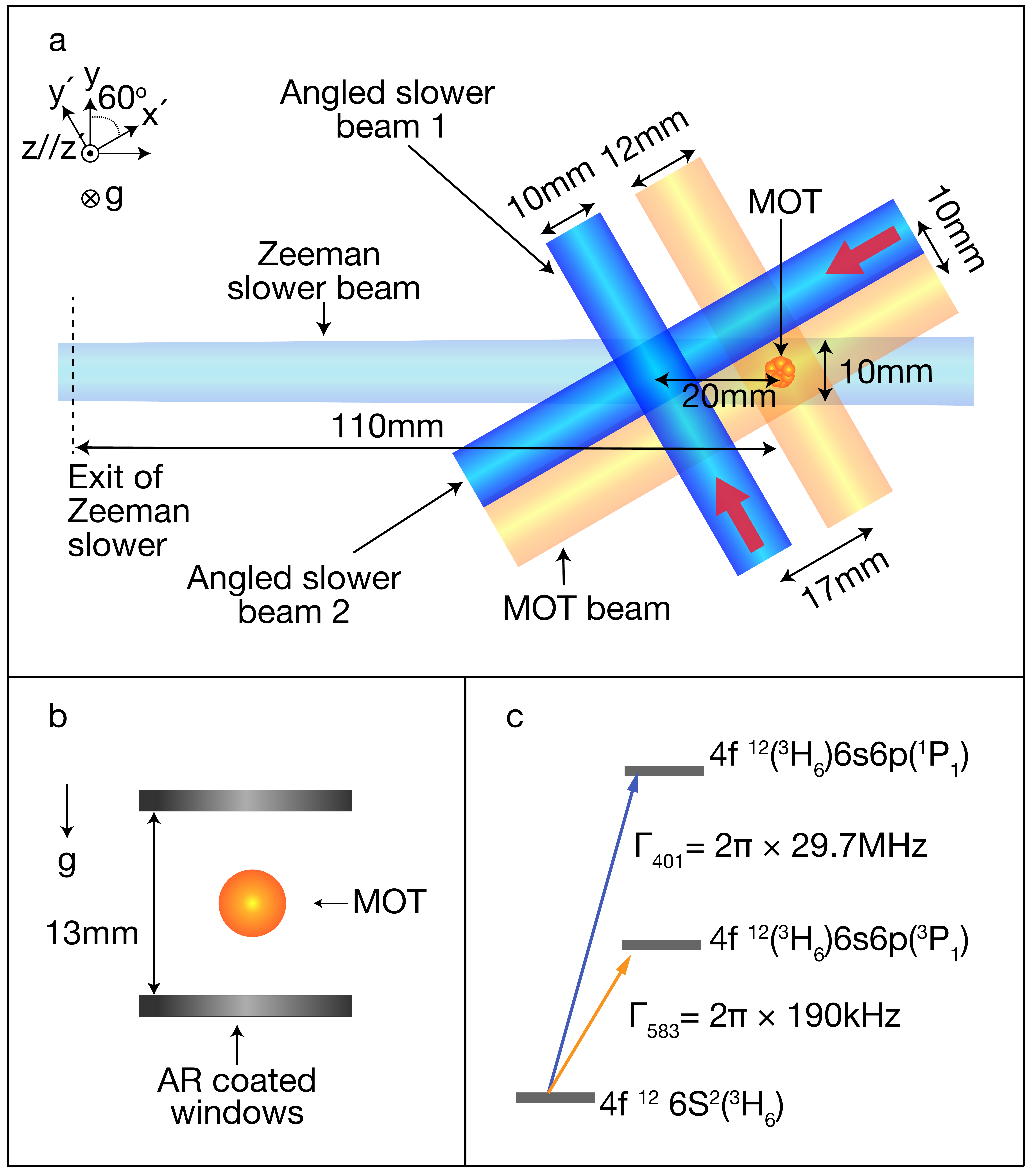}
	\caption{ {\bf Schematics of two-stage slowing, the science chamber, and the level diagram} (a) Layout of two-stage slowing scheme near the narrow-line MOT. Two 401~nm beams intersect 20~mm ahead of the MOT position, wherein atoms exiting from the Zeeman slower are further decelerated along the x direction and slowed along the y direction. (b) View of the narrow-line MOT and the recessed viewports. In the science chamber, the diameter of the MOT beam is fully maximized up to 12~mm. The gap between two recessed viewports is 13~mm. (c) Broad singlet and narrow triplet transitions are used for two-stage slowing and narrow-line MOT loading. }
	\label{fig2}
\end{figure}

%\paragraph*{\bf  Description of $^{166}$Er MOT} 
In the chamber, a narrow-line MOT is formed by three pairs of 50~mW retro-reflected 583~nm laser beams, resulting in a saturation parameter of $s\approx340$ ($I_{s,583}=0.13$ mW/cm$^2$). During the loading process, MOT beams are detuned by -21 $\Gamma_{583}$ from the narrow transition while using a magnetic field gradient of 3.5~G/cm along the quadrupole axis. To increase the MOT capture velocity, we broaden the laser linewidth up to 2.0~MHz with an acousto-optic modulator (AOM) being frequency modulated at 200~kHz. The science chamber is equipped with recessed viewports for a high-resolution imaging setup (see Fig.~\ref{fig1}) and the narrow-line MOT is formed at the center of the chamber without noticeable gravitational sag. %In this configuration, we trap $\sim$1$\times 10^7$ $^{166}$Er atoms without second-stage slowing beams. 
Note that we optimize the Zeeman slower at low optical power and therefore minimize possible off-resonant optical pumping caused by the slower light, which potentially reduces the loading efficiency into the MOT.

\section{Characterization of two-stage slowing}

\begin{figure}[tbp]
	\includegraphics[width=1.03\linewidth]{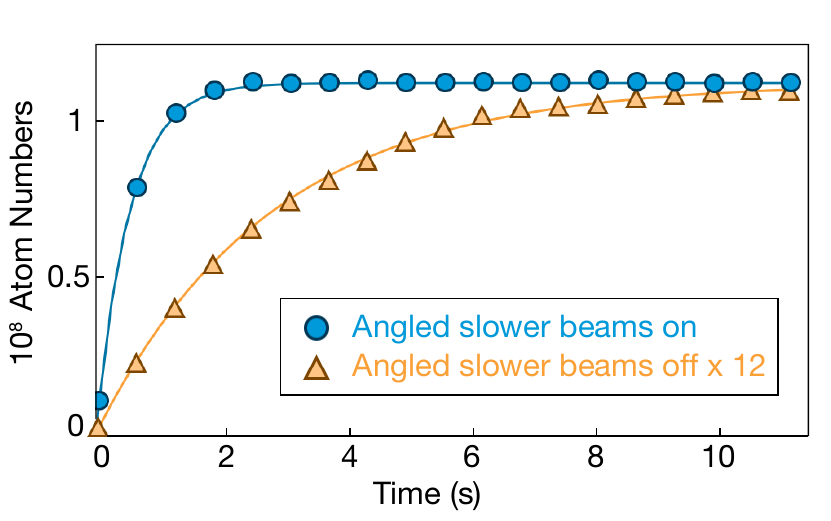}
	\caption{\textbf{Enhanced $^{166}$Er MOT loading with angled slower beams}  Atom number $N_{\text MOT}$ loaded in the narrow-line MOT is measured by atomic fluorescence (and absorption imaging) as a function of time at the oven temperature of 1200$^{\circ}$C. The second-stage slowing significantly improves the steady-state atom number by more than an order of magnitude. The Zeeman slower magnetic field is kept the same for both measurements. The data without angled slower beams is multiplied by 12 to simplify comparison. The blue circle (yellow triangle)  is a fit result with the empirical equation $N_{\text MOT}(t)=R\tau(1-e^{-t/\tau})$ where $R$ is the MOT loading rate and $\tau^{-1}$ the loss rate from the MOT. The typical MOT loading time ($\tau$) is about 0.5~s with the angled slower beams. }
	\label{fig3}
\end{figure}

To increase the MOT loading flux, we implement  second-stage slowing with two angled beams being circularly-polarized and red-detuned by 34 MHz (-1.13 $\Gamma_{401}$). Those 401~nm beams intersect 20~mm away from the narrow-line MOT position, as depicted in Fig.~\ref{fig2}(a). Before entering the science chamber, atoms are slowed by the conventional Zeeman slower along the longitudinal direction (x direction in Fig.~\ref{fig2}(a)). Inside the chamber, angled slower beams further decelerate the slow atoms exiting from the Zeeman slower along both the longitudinal and transverse directions, and effectively reduce the transverse velocity distribution at the MOT position, increasing the MOT population significantly (see Fig.~\ref{fig3}). The angled slowing beams are generated from an injection-locked slave laser, resulting in 21~mW and 7~mW of power for beam 1 and 2, respectively. We optimize the power distribution and detuning of these two beams based on the MOT atom number. The net scattering force along the y direction is balanced by the power imbalance of the angled slower beams.

In Fig.~\ref{fig3}, we monitor the trapped atom number in a $^{166}$Er MOT with and without the angled slower beams at an oven temperature of 1200$^{\circ}$C. We observe that the MOT loading rate is significantly enhanced (by more than one order of magnitude), from $R=3.5 \times 10^6 s^{-1}$ to $R=2.3 \times 10^8 s^{-1}$, which produces the saturated atom number of $1.1 \times 10^8$ within two seconds. The atom number is consistently calibrated both by fluorescence detection and absorption imaging. Without the angled slowing beams, the total MOT atom number is around $\sim 3 \times 10^7$ with experimental parameters (e.g. Zeeman slower magnetic field, Zeeman slower beam detuning, MOT quadrupole magnetic field, and MOT beam detuning) being separately optimized. It is worthwhile to note that with angled slower beams the saturated atom number in the MOT increases up to $2\times 10^8$ if the oven temperature is increased to 1250$^{\circ}$C. The two-stage slowing scheme also works for other isotopes such as $^{168}$Er with similar performance.

\begin{figure}[tbp]
	\includegraphics[width=1.0\linewidth]{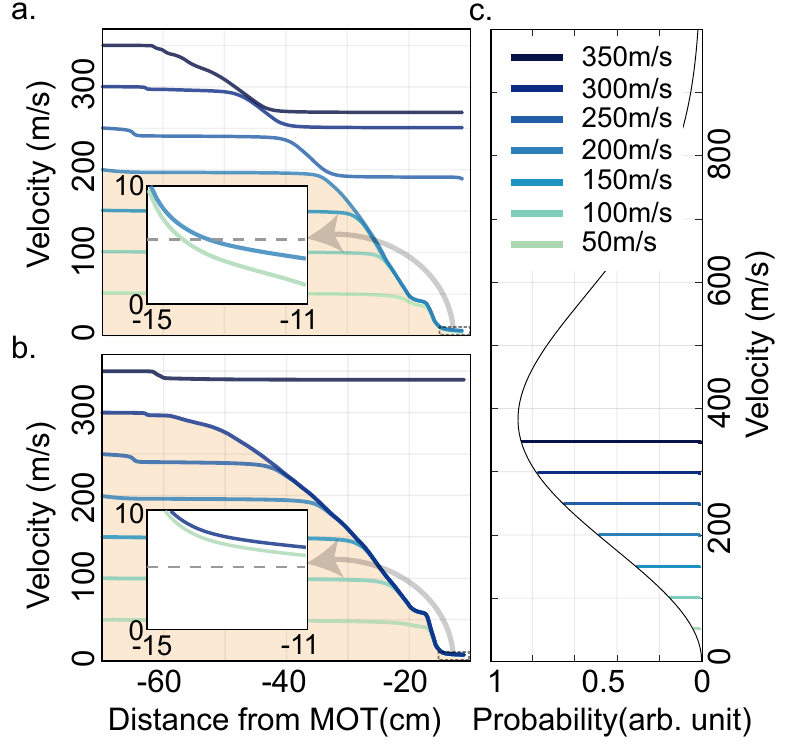}
	\caption{\textbf{Enhanced capture velocity of Zeeman slower with two-stage slowing } Velocity of Er atoms along the Zeeman slower with angled slower beams off (a) and on (b). The magnetic field in the Zeeman slower is separately optimized on the MOT loading in each case (a,b). For example, the final velocities of Er atoms with an initial longitudinal speed of 200~m/s are 3.8~m/s and 7.6~m/s after the Zeeman slower in (a) and (b), respectively. The insets show zooms of the velocity in the range between -15~cm and -11~cm. The MOT is located at 0~cm. The shade indicates the capture velocity of the Zeeman slower in (a,b).(c) Maxwell-Boltzmann distribution of Er atoms at 1200$^{\circ}$C. Colored lines show different initial velocities. }
	\label{fig4}
\end{figure}

We  partially attribute the enhanced atom number to the improved capture velocity of Zeeman slower as described in Fig.~\ref{fig4}. In Fig.~\ref{fig4}(a,b), we show the numerically calculated trajectories of atoms with various initial velocities from the effusive oven to the MOT with and without the angled slowing beams, respectively. Taking a sample of atoms with a given velocity distribution, a full trajectory is simulated by calculating the velocity and position change of atoms during small time intervals. For each case,  experimental parameters being employed in our experiment are used for the simulation. It suggests that the effective capture velocity of the Zeeman slower is enhanced by angled slowing beams. A detailed theoretical analysis can be an interesting direction for future work.

\section{Conclusion}

In conclusion, we have demonstrated a simple and cost-effective method to produce a large number of Er atoms in a narrow-line MOT. We apply a two-stage slowing scheme, for the first time in an Er experiment, which significantly enhances the loading efficiency into the MOT. Our experiment suggests that the near-resonant angled slowing beams  in front of the MOT further slow down atoms exiting from the Zeeman slower and therefore enhance the loading rate of the MOT. This method can be easily applied to laser cooling and trapping other similar elements with high melting points.  Therefore, our method may not only greatly simplify the current approach for producing ultracold magnetic atoms, but also pave the way for the broad search of  quantum degenerate gases of elements that have never been explored as far.

\paragraph*{\bf Acknowledgement} We thank Y.M. Tso and D. Lee for their initial contribution. G.-B.J. and S.D acknowledge the  support from the Collaborative Research Fund (C6005-17G) of the Hong Kong Research Grants Council (RGC). G.-B.J. acknowledges additional supports from the RGC and the Croucher Foundation through 16311516, 16305317, 16304918, N-HKUST601/17. P.C. acknowledges the support by Shanghai Natural Science Foundation (Grant No.18ZR1443800) and
Innovation Promotion Association of the Chinese Academy of Science. M.H. acknowledges the support from the Huawei PhD scholarship. W.Y is supported by UROP program at HKUST.

%\bibliography{allref.bib}

%========================================================================================================

\end{document}